\documentclass[aps,twocolumn,showpacs]{revtex4}

\usepackage[xdvi]{graphicx}
\usepackage{latexsym}
\usepackage{amsbsy}
\usepackage{amssymb}
\usepackage{amsmath}

\begin{document}
\date{\today}

\title{Comparison of quantum and classical relaxation in spin dynamics} 

\author{R.\ Wieser} 

\affiliation{Institut f\"ur Angewandte Physik,
    Universit\"at Hamburg, D-20355 Hamburg, Germany
}

\begin{abstract}
The classical Landau-Lifshitz equation with damping term has been derived from 
the time evolution of a quantum mechanical wave function under the assumption 
of a non-hermitian Hamilton operator. Further, the trajectory of a classical 
spin $\mathrm{S}$ has been compared with the expectation value of the spin 
operator $\mathrm{\hat{S}}$. A good agreement between classical and quantum 
mechanical trajectories can be found for Hamiltonians linear in 
$\mathrm{\hat{S}}$ respectively $\mathrm{S}$. Quadratic or higher order terms 
in the Hamiltonian result in a disagreement.
 
\end{abstract}

\maketitle

The Landau-Lifshitz equation \cite{landauPZS35} is one of the most often used 
equations in physics. This equation is of importance not only in micromagnetism 
\cite{brownBOOK63} or for the spin dynamics at the atomic level 
\cite{antropovPRL95ETAL}, but also in disciplines like astronomy 
\cite{boernerAA75ETAL}, biology \cite{bellPRE07ETAL}, chemistry 
\cite{huchtEPL07ETAL}, and medicine \cite{witteJSMN12ETAL}. In micromagnetism 
it describes the motion of a magnetic moment in a local magnetic field . The 
equation of motion can be augmented easily by additional interactions that can 
be incorporated into an effective field or, e.g., by temperature effects. 
Moreover, there are many similar, equivalent, or alternative approaches namely 
the Bloch equation \cite{blochPR46}, the Ishimori equation \cite{ishimoriPTP84} 
and the Landau-Lifshitz-Bloch equation \cite{garaninPRB97}. All these 
approaches are capable to describe magnetization dynamics starting from a 
single atomic spin up to several micrometers. 

Originally, the Landau-Lifshitz equation was introduced as a pure 
phenomenological equation \cite{landauPZS35}. Later it has been shown that the 
precessional term can be derived by quantum mechanics \cite{wieserPRB11}, but 
the damping term in the Landau-Lifshitz equation remained phenomenological 
untill Gilbert proposed to use the Lagrange formalism with the classical 
Rayleigh damping instead of the original Landau-Lifshitz damping to improve the 
equation, resulting in the Landau-Lifshitz-Gilbert equation 
\cite{gilbertIEEE04}. 

In this publication I will describe an alternative and thereby closing the lack 
of knowledge: a simple derivation of the Landau-Lifshitz equation with damping 
starting from the quantum mechanical time evolution will be given. Such a 
derivation provides a deeper understanding of the underlying mathematics and 
the connection between quantum mechanics and classical physics. 

The derivation starts with the quantum mechanical time evolution of the state 
$|\psi(t)\rangle$:
\begin{equation} \label{psiTIME}
|\psi (t+\Delta t)\rangle = \hat{U}\left(t+\Delta t,t\right)|\psi(t)
\rangle \;.  
\end{equation}
Under the assumption of a small time step $\Delta t$ we can expand the time 
evolution operator $\hat{U}\left(t+\Delta t,t\right) \approx 
\left(\hat{1} -\mathrm{i}\hat{{\cal H}} \Delta t/\hbar\right) + 
{\cal O}\left(\Delta t^2 \right)$:
\begin{equation} \label{psi1}
|\psi^{(1)}(t+\Delta t)\rangle \approx \left(\hat{1} 
-\frac{\mathrm{i}\hat{{\cal H}}\Delta t}{\hbar}\right)|\psi(t)\rangle  
\end{equation}
If we further assume that we have a non-Hermitian Hamilton operator:
$\hat{{\cal H}}^+ \neq \hat{{\cal H}}$ we get:  
\begin{equation}
\langle\psi^{(1)}(t+\Delta t)| \approx \langle\psi(t)|\left(1 +
\frac{\mathrm{i} \hat{{\cal H}}^+\Delta t}{\hbar}\right) 
\end{equation}
and the norm: 
\begin{eqnarray} \label{norm}
n^2 &=& \langle \psi^{(1)}(t+\Delta t)|\psi^{(1)}(t+\Delta t)\rangle 
\nonumber \\
&=& \langle\psi(t)|\left(1 +\frac{\mathrm{i}\hat{{\cal H}}^+\Delta t}{\hbar}
\right)\left(1 - \frac{\mathrm{i}\hat{{\cal H}}\Delta t}{\hbar}\right)|
\psi(t)\rangle \nonumber \\
&=& 1 - \frac{\mathrm{i}}{\hbar}\Delta t \langle\psi(t)| \hat{{\cal H}} -  
\hat{{\cal H}}^+ |\psi(t)\rangle = 1 - r\;.
\end{eqnarray}
Now we are now looking for a normalized wave function and make the ansatz:
\begin{equation} \label{ansatz}
|\psi(t+\Delta t)\rangle = \frac{|\psi^{(1)}(t+\Delta t)\rangle}{
\sqrt{1-r}}
\end{equation}
Then, Eq.~(\ref{psi1}) can be rewritten
\begin{equation}
\frac{|\psi^{(1)}(t+\Delta t)\rangle - |\psi(t)\rangle}{
\Delta t} = -\frac{\mathrm{i}}{\hbar}\hat{{\cal H}}|\psi(t)\rangle\;, 
\end{equation}
and with Eq.~(\ref{ansatz})
\begin{equation}
\frac{|\psi(t+\Delta t)\rangle \sqrt{1-r} - |\psi(t)\rangle}
{\Delta t} = -\frac{\mathrm{i}}{\hbar} \hat{{\cal H}}|\psi(t)\rangle\;.
\end{equation}
Further, with the Taylor expansion: $\sqrt{1-r} \approx 1 -\frac{1}{2}r$ we 
get: 
\begin{equation}
\frac{|\psi(t+\Delta t)\rangle - |\psi(t)\rangle}{\Delta t} - 
\frac{1}{2}\frac{r}{\Delta t}|\psi(t+\Delta t)\rangle  = 
-\frac{\mathrm{i}}{\hbar}\hat{{\cal H}}|\psi(t)\rangle  
\end{equation}
where $r$ is given by Eq.~(\ref{norm}). In the limit $\Delta t 
\rightarrow 0$ the differential quotient becomes a differential operator 
$\mathrm{d}t$ and $|\psi(t+\Delta t)\rangle$ becomes $|\psi(t)\rangle$.
Finally, we get the following modified time dependent Schr{\"o}dinger equation:
\begin{equation} \label{molmerEq}
\mathrm{i}\hbar \frac{\mathrm{d}}{\mathrm{d}t}|\psi(t)\rangle = 
\left(\hat{{\cal H}} + \langle\psi(t)|\frac{\hat{{\cal H}}^+ - 
\hat{{\cal H}}}{2}|\psi(t)\rangle\right)|\psi(t)
\rangle \;.
\end{equation} 
This formula is identical with the equation proposed by K. M{\o}lmer et al. 
\cite{molmerJOSAB93ETAL} for the calculation of Monte Carlo wave functions in 
quantum optics.
 
With $\hat{{\cal H}} = \hat{\mathrm{H}} -\mathrm{i}\lambda \hat{\Gamma}$, 
$\hat{{\cal H}}^+ = \hat{\mathrm{H}} +\mathrm{i}\lambda \hat{\Gamma}$ 
($\lambda \in \mathbb{R}_0^+$, $ \hat{\Gamma}$ hermitian), and 
$\langle\hat{\Gamma} \rangle = \langle\psi(t)|\hat{\Gamma}|\psi(t) \rangle$ 
Eq.~(\ref{molmerEq}) becomes:
\begin{equation} \label{molmerEq2}
\mathrm{i}\hbar \frac{\mathrm{d}}{\mathrm{d}t}|\psi(t)\rangle = 
(\hat{\mathrm{H}} -\mathrm{i}\lambda [\hat{\Gamma} - \langle\hat{\Gamma} 
\rangle])|\psi(t)\rangle
\end{equation} 
N. Gisin \cite{gisinHelvPhysActa81} has proposed a similar equation, however, 
with the use of $\hat{{\cal H}} = \hat{\mathrm{H}} -\mathrm{i}\lambda 
\hat{\mathrm{H}}$:
\begin{equation} \label{Gisin}
\mathrm{i} \hbar \frac{\mathrm{d}}{\mathrm{d}t}|\psi(t)\rangle = 
(\hat{\mathrm{H}} -\mathrm{i}\lambda [\hat{\mathrm{H}} - 
\langle \hat{\mathrm{H}} \rangle])| \psi(t)\rangle
\end{equation}
This special case of Eq.~(\ref{molmerEq2}) is the quantum mechanical 
counterpart of the Landau-Lifshitz equation as will be shown below. 

Eq.~(\ref{Gisin}) can be rewritten as:
\begin{equation} \label{GisinII}
\mathrm{i}\hbar \frac{\mathrm{d}}{\mathrm{d}t}|\psi \rangle =  
\Big(\hat{\mathrm{H}} - \mathrm{i}\lambda \Big[\hat{\mathrm{H}}; 
|\psi \rangle   \langle \psi |\Big]\Big)|\psi \rangle
\end{equation}
and for the corresponding transposed equation we can use the fact that 
$\hat{\mathrm{H}}^T = \hat{\mathrm{H}}^+ = \hat{\mathrm{H}}$. Then, the 
transposed commutator is given by
\begin{equation} \label{Transcommutator2}
\Big[\hat{\mathrm{H}}; |\psi \rangle   \langle \psi |\Big]^T = 
-\Big[\hat{\mathrm{H}}; |\psi
 \rangle   \langle \psi |\Big]\;,
\end{equation}
and therefore the corresponding transposed equation:
\begin{equation} \label{GisinIV}
-\mathrm{i}\hbar \frac{\mathrm{d}}{\mathrm{d}t} \langle \psi | =  \langle \psi 
| \Big(\hat{\mathrm{H}} -\mathrm{i}\lambda \Big[\hat{\mathrm{H}}; |\psi 
\rangle  \langle \psi |\Big] \Big)
\end{equation}
Now, we are able to write down the corresponding von Neumann or quantum 
Liouville equation \cite{garaninAdvChemPhys11} of the density operator 
$\hat{\rho}$: 
\begin{eqnarray} 
\frac{\mathrm{d}\hat{\rho}}{\mathrm{d}t} = \frac{\mathrm{d}}{\mathrm{d}t} 
\Big(|\psi \rangle \langle \psi |\Big)  &=& 
\frac{\mathrm{d} |\psi \rangle}{\mathrm{d}t} \langle \psi | +
 |\psi \rangle \frac{\mathrm{d} \langle \psi |}{\mathrm{d}t} \nonumber \\
&=& \frac{\mathrm{i}}{\hbar} \Big[\hat{\rho}; \hat{\mathrm{H}} \Big] - 
\frac{\lambda}{\hbar} \Big[\hat{\rho};\Big[\hat{\rho}; \hat{\mathrm{H}} \Big]
\Big]\;. 
\end{eqnarray}
In the Schr{\"o}dinger picture the time dependence of the expectation value 
$\langle \mathbf{\hat{S}} \rangle$ is invested 
in $\hat{\rho}$ and in the Heisenberg picture in $\mathbf{\hat{S}}$, which is 
in the Schr{\"o}dinger picture time independent:
\begin{eqnarray}
\mathrm{i}\hbar \frac{\mathrm{d} \langle \mathbf{\hat{S}} \rangle}{\mathrm{d}t} 
&=& \mathrm{i} \hbar \mathrm{Tr} \left(\frac{\mathrm{d}\hat{\rho}}{\mathrm{d}t} 
\mathbf{\hat{S}} \right) 
\end{eqnarray}
and therefore we find under usage of the cyclic change under the trace:  
\begin{equation} \label{SErwartungswertSPicture}
\frac{\mathrm{d} \langle \mathbf{\hat{S}} \rangle}{\mathrm{d}t} = 
-\frac{\mathrm{i}}{\hbar}\left\langle \Big[\mathbf{\hat{S}} ;\hat{\mathrm{H}} 
\Big] \right\rangle + \frac{\lambda}{\hbar} \left\langle \Big[\mathbf{\hat{S}} 
; \Big[\hat{\rho}; \hat{\mathrm{H}}\Big]\Big]\right\rangle
\end{equation}
Please notice there is still a $\hat{\rho}$ included on the right hand side of 
Eq.~(\ref{SErwartungswertSPicture}).\\
To get the Heisenberg equation we interpret the operators in the Heisenberg 
picture and skip the bra's $\langle \psi |$ and ket's $| \psi \rangle$ on both 
sides of the equation ($\langle \mathbf{\hat{S}} \rangle = \langle \psi 
|\mathbf{\hat{S}}| \psi \rangle$). The expectation values are identical in 
both pictures and therefore we finally get: 
\begin{equation} \label{HeisenberEOM}
\frac{\mathrm{d} \mathbf{\hat{S}}}{\mathrm{d}t} = -\frac{\mathrm{i}}{\hbar}\Big[
\mathbf{\hat{S}} ;\hat{\mathrm{H}} \Big]  + \frac{\lambda}{\hbar} \Big[\mathbf{
\hat{S}} ; \Big[\hat{\rho}; \hat{\mathrm{H}} \Big]\Big]
\end{equation}
The problem is, there is still an additional $\hat{\rho}$ instead of 
$\mathbf{\hat{S}}$. 
For $S = \frac{1}{2}$ the density matrix is given by:
\begin{equation} \label{DensityMatrixS12}
\hat{\rho} =  \frac{1}{2}\left(\mathbf{\hat{1}} + \langle \mathbf{\hat{\sigma}} 
\rangle \mathbf{\hat{\sigma}} \right) \;.
\end{equation}
The factor $\frac{1}{2}$ is just for the normation because $\mathrm{Tr}
\hat{\rho} = 1$. The unity matrix $\mathbf{\hat{1}}$ does commutate with 
$\hat{\mathrm{H}}$ therefore this term can be skipped and $\hat{\rho}$ is 
equal to the polarization $\langle \mathbf{\hat{\sigma}} \rangle 
\mathbf{\hat{\sigma}}$ with the Pauli matrix vector $\mathbf{\hat{\sigma}} = 
(\hat{\sigma}_x,\hat{\sigma}_y,\hat{\sigma}_z)$. Here, $\sigma_\eta$, $\eta \in 
\{x,y,z\}$ are the Pauli matrices. For general $S$ the polarization $\langle 
\mathbf{\hat{\sigma}} \rangle \mathbf{\hat{\sigma}}$ has to be replaced by 
$\langle \mathbf{\hat{S}} \rangle \mathbf{\hat{S}}/S$ with the corresponding 
spin matrix vector $\mathbf{\hat{S}} = (\hat{S}_x,\hat{S}_y,\hat{S}_z)$ and 
$\hat{S}_\eta = \bigotimes\limits_{n=1}^{2S} \sigma_n^\eta$, $(\eta \in \{x,y,z\})$ 
\cite{fanoRMP57}:
\begin{equation} \label{DensityMatrixS}
\hat{\rho} \approx \frac{\langle \mathbf{\hat{S}} \rangle 
\mathbf{\hat{S}}}{\hbar S} \;.
\end{equation}
Under the assumption of being in a pure state: $|\langle \mathbf{\hat{S}} 
\rangle| = S$, and the further assumption that the $\mathbf{\hat{z}}$-axis is 
the quantization axis: 
$\langle \mathbf{\hat{S}} \rangle = \langle \hat{S}_z \rangle = \hbar S$ we get:
\begin{equation} \label{DensityMatrixSII}
\hat{\rho} \approx \frac{\langle \hat{S}_z \rangle \mathbf{\hat{S}}}{\hbar S} = 
\mathbf{\hat{S}}\;.
\end{equation}
Putting this in Eq.~(\ref{HeisenberEOM}) gives the Heisenberg equation:
\begin{equation} \label{HeisenberEOMIII}
\frac{\mathrm{d} \mathbf{\hat{S}}}{\mathrm{d}t} = -\frac{\mathrm{i}}{\hbar}
\Big[\mathbf{\hat{S}} ;\hat{\mathrm{H}} \Big]  + \frac{\lambda}{\hbar} 
\Big[\mathbf{\hat{S}} ; 
\Big[\mathbf{\hat{S}}; \hat{\mathrm{H}} \Big]\Big] \;.
\end{equation}
In a previous publication \cite{wieserPRB11} I have shown that 
\begin{equation}
\frac{\mathrm{i}}{\hbar}\Big[\mathbf{\hat{S}} ;\hat{\mathrm{H}} \Big] = 
\mathbf{\hat{S}} \times \frac{\partial\hat{\mathrm{H}}}{\partial{\mathbf S}} + 
{\cal O}(\hbar)\;,
\end{equation}
where the cross product and the gradient directly follow from the 
definition of the commutator \cite{lakshmananPTRS11} and the additional term 
occurs if the Hamilton operator is not linear in $\mathbf{\hat{S}}_n$. The 
double commutator term on the right hand side is more complicated. Here, we 
have to know that within the Clifford Algebra
$\mathbf{\hat{S}} \times \mathbf{\hat{S}} =\mathrm{i}\mathbf{\hat{S}}$ and 
$(\mathbf{\hat{S}} \times \mathbf{\hat{S}}) \times \mathbf{\hat{H}} = 
\mathbf{\hat{S}} \times (\mathbf{\hat{S}} \times \mathbf{\hat{H}})$ holds, 
which is not the case for normal vectors. Here $\mathbf{\hat{S}} = 
(\hat{S}_x,\hat{S}_y,\hat{S}_z)$ and $\mathbf{\hat{H}} = 
(\hat{H}_x,\hat{H}_y,\hat{H}_z)$ are matrix vectors. Alternatively, we can 
use the following relation of the $SO(3)$ Lie algebra:
\begin{equation}
\mathbf{x} \times \mathbf{y} = \mathbf{\hat{x}}\mathbf{y} \equiv 
\mathbf{\hat{x}}\mathbf{\hat{y}} - \mathbf{\hat{y}}\mathbf{\hat{x}}
= [\mathbf{\hat{x}},\mathbf{\hat{y}}]
\end{equation}
where $\mathbf{x}$ and $\mathbf{y}$ are normal vectors and 
$\mathbf{\hat{x}}$ and $\mathbf{\hat{y}}$ 
are 3x3 skew-symmetric matrices:
\begin{equation}
\mathbf{x} = \left(\begin{array}{c} x_1 \\ x_2 \\ x_3 \end{array}\right)\,, 
\hspace{3mm} \mathbf{\hat{x}} = \left(\begin{array}{ccc} 0 & -x_3 & x_2 \\ x_3 &
0 & -x_1 \\ -x_2 & x_1 & 0 \end{array} \right)\,, 
\end{equation}
with $\mathbf{y}$ and $\mathbf{\hat{y}}$ accordingly. This relation can be 
proven with the aid of the Jacobi identity of the cross product. For details 
and additional depictions see \cite{liuJMST09}.

In the limit $S \rightarrow \infty$ and $\hbar \rightarrow 0$ we get the 
classical Landau-Lifshitz equation: 
\begin{eqnarray}
  \label{llgR}
  \frac{\mathrm{d}{\mathbf S}}{\mathrm{d} t} = -
  \frac{\gamma}{\mu_S}{\mathbf S}\times {\mathbf H}_{\mathrm{eff}} +
  \frac{\lambda}{\mu_S}{\mathbf S}\times \left({\mathbf S}
 \times {\mathbf H}_{\mathrm{eff}} \right) \,.  
\end{eqnarray}
$\gamma$ is the gyromagnetic ratio coming from the relation between magnetic 
moment ${\boldsymbol \mu}$ and spin, $\mu_S = |{\boldsymbol \mu}|$ comes from 
the normalization, $\lambda$ the damping constant, and 
${\mathbf H}_{\mathrm{eff}} = -\partial\mathrm{H}/\partial{\mathbf S}$ 
the effective field, with classical Hamilton function $\mathrm{H}$.  

\begin{figure}[h]
\begin{center}
  \includegraphics*[bb = 157 545 458 752, width=6.5cm]{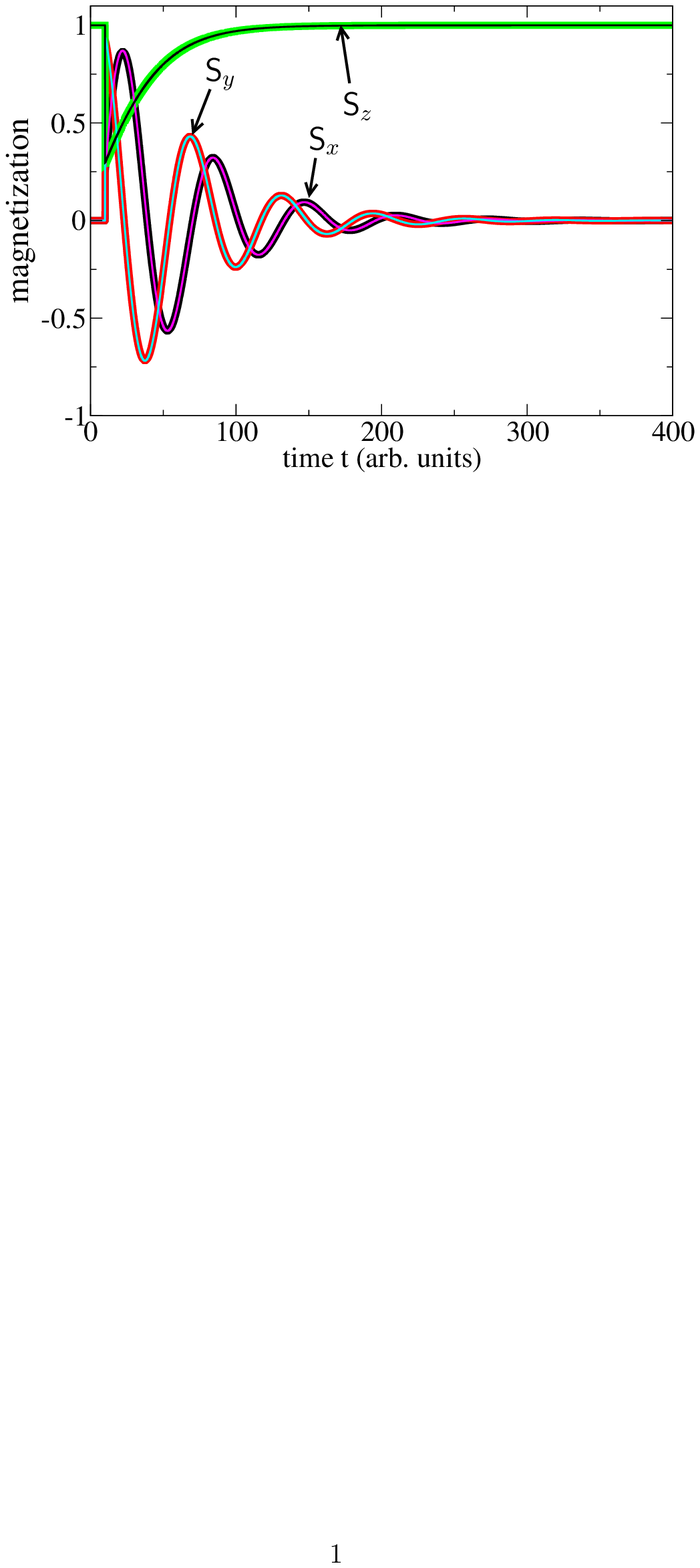}
  \end{center}
  \caption {(color online) Magnetization as function of time after a gaussian 
field pulse: comparison of classical (bold lines) and quantum mechanical 
trajectories (thin lines).\\
($D_z = 0$, $\mu_SB_z = 0.1$, $T_W = 0.02$, $t_0 = 10$, $\mu_SB_0^x = 25.27$, and
$\lambda = 0.2$)}     
  \label{f:pic1}
\end{figure}

\begin{figure}[h]
\begin{center}
  \includegraphics*[bb = 157 532 458 752, width=6.5cm]{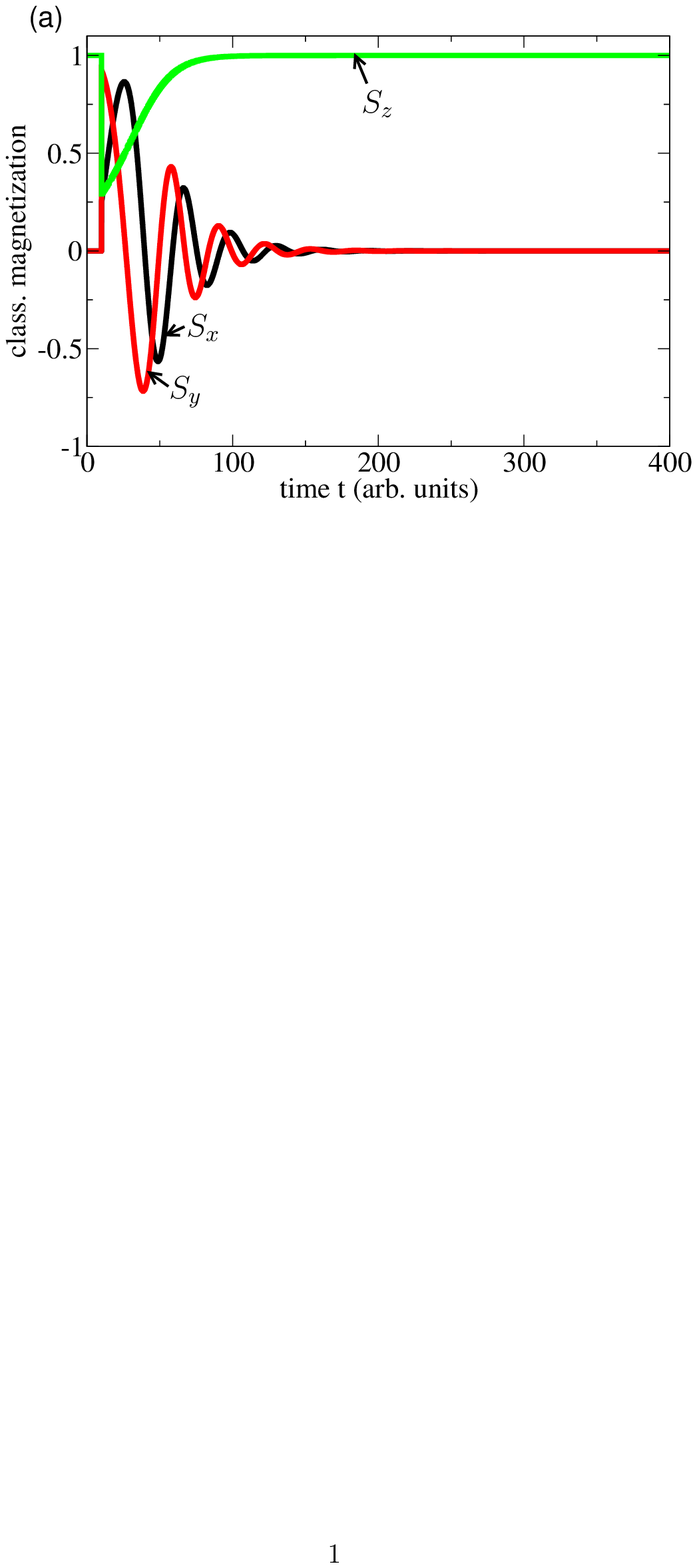}
  \includegraphics*[bb = 157 532 458 752, width=6.5cm]{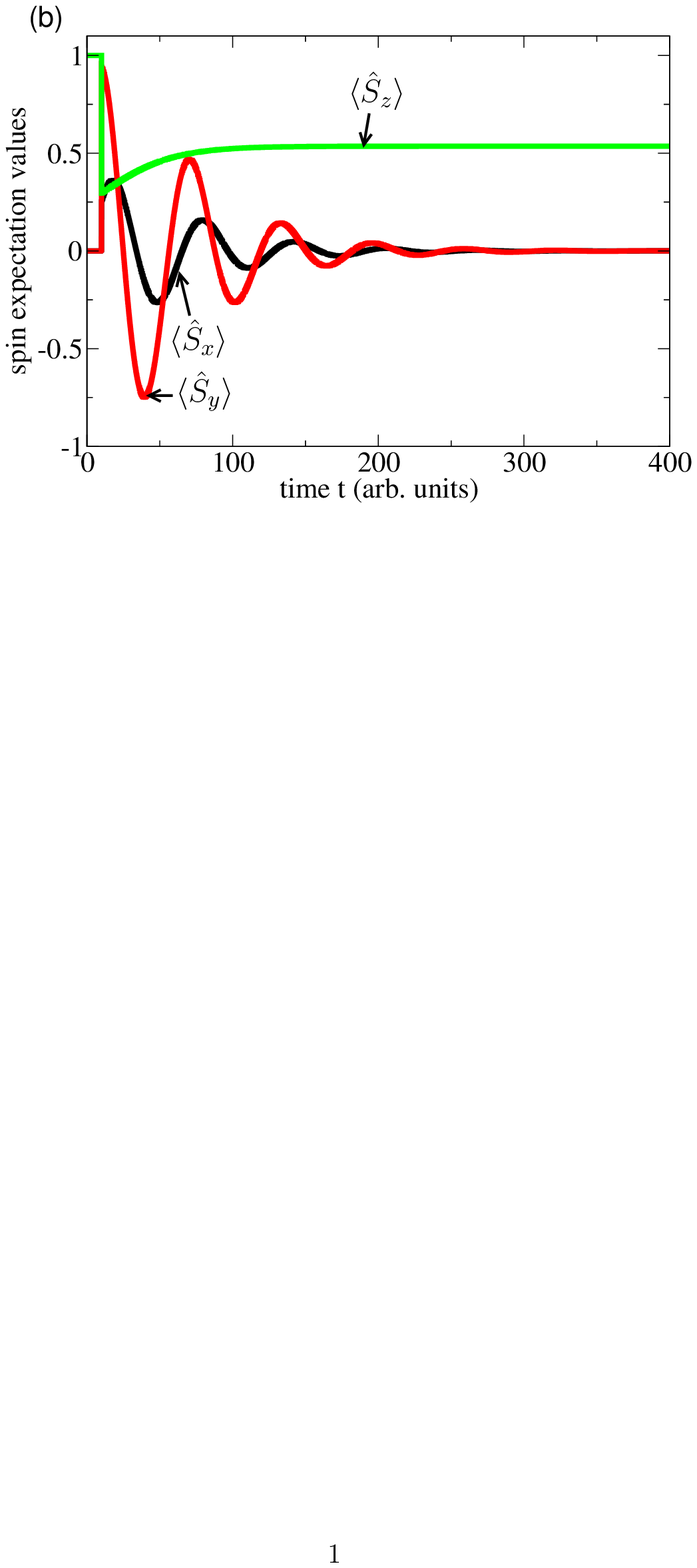}
  \end{center}
  \caption {(color online) Magnetization as function of time after a gaussian 
field pulse: (a) classical trajectory $S_\eta$, (b) quantum mechanical 
expectation values $\langle \hat{S}_\eta \rangle$, $\eta \in\{x,y,z\}$ \\
($D_z = 0.1$, $\mu_SB_z = 0$, $T_W = 0.02$, $t_0 = 10$, $\mu_SB_0^x = 25.27$, and
$\lambda = 0.2$)}     
  \label{f:pic2}
\end{figure}

To prove the agreement between the TDSE and the Landau-Lifshitz Eq. we have 
performed numerical calculations. In the following we use a simple single spin 
model. The description is just an example and can be extended to systems with 
$N>1$. The corresponding Hamilton operator $\hat{\mathrm{H}}$ is given by: 
\begin{eqnarray}
\hat{\mathrm{H}} = - D_z \left(\hat{S}_z\right)^2  -\mu_SB_z \hat{S}_z
-\mu_SB_x(t)\hat{S}_x 
\end{eqnarray} 

The first term of the Hamiltonian describes a uniaxial anisotropy with the
$z$-axis as the easy axis. The second term represents a static external
magnetic field in $+z$-direction. The last term is a time-dependent
field pulse 
\begin{eqnarray}
B_x(t) = B_0^x{\mathrm e}^{-\frac{1}{2}\left(\frac{t-t_0}{T_W}\right)^2}
\end{eqnarray} 
with gaussian shape to excite the spin. In an experimental setup using single 
atoms such an excitation can be realized, e.g., by a current pulse coming from 
an STM (scanning tunneling microscope) tip. In the following we investigate 
the two situations: either (i) $D_z = 0$ and $B_z \neq 0$ or vice versa (ii) 
$D_z \neq 0$ and $B_z = 0$. In the case (i) all terms of the Hamiltonian are 
linear in $\hat{\mathbf{S}}$. In the second case (ii) the Hamiltonian
contains a quadratic term. 

In a previous publication \cite{wieserPRB11} we have shown that in case (i) 
under the assumption of a negligible damping $(\lambda = 0)$ a good agreement 
between classical and quantum spin dynamics can be obtained. Case (ii) shows 
without relaxation a disagreement between classical and quantum spin dynamics 
due to the noncommutativity of the quadratic terms. 

Fig.~\ref{f:pic1} shows the results of the calculation of a single spin with
relaxation for the case (i) (without anisotropy). Again an excellent agreement 
between  quantum mechanical expectation values $\langle \hat{S}_\eta \rangle$ 
$\eta \in\{x,y,z\}$ and the classical trajectories $S_\eta$ of Landau-Lifshitz 
Eq. is found. In case (ii) [Fig.~\ref{f:pic2}] the classical trajectories (a) 
and quantum mechanical expectation values (b) disagree. This is caused by the 
quadratic Hamilton operator due to the anisotropy \cite{wieserPRB11}. In this 
case the commutator $[\mathbf{\hat{S}}_n ;\hat{\mathrm{H}}]$ leads to 
$\mathbf{\hat{S}} \times \partial\hat{\mathrm{H}}/\partial{\mathbf S}$ plus an 
additional term of the order of $\hbar$. This additional term vanishes in the 
classical limit $S\rightarrow \infty$ and leads to the disagreement between 
quantum mechanical and classical trajectory. In the case of a linear 
Hamiltonian the commutator $[\mathbf{\hat{S}}_n ; \hat{\mathrm{H}}]$ does not 
lead to an additional correction and the both trajectories show a perfect 
agreement. Therefore, we can say that the second commutator in the damping term 
does not produce any corrections.    
 
\begin{figure}[h]
\begin{center}
  \includegraphics*[bb = 157 524 458 752, width=6.5cm]{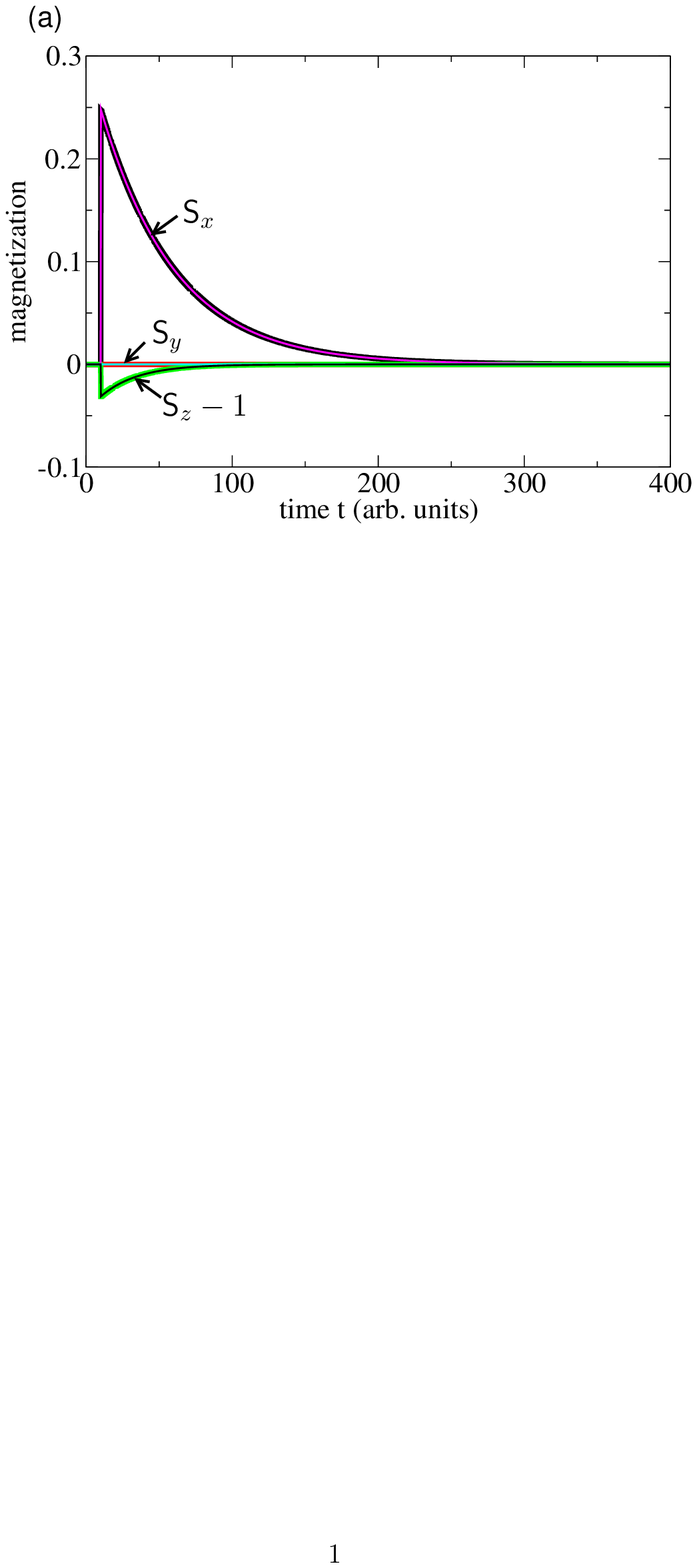}
  \includegraphics*[bb = 157 524 458 752, width=6.5cm]{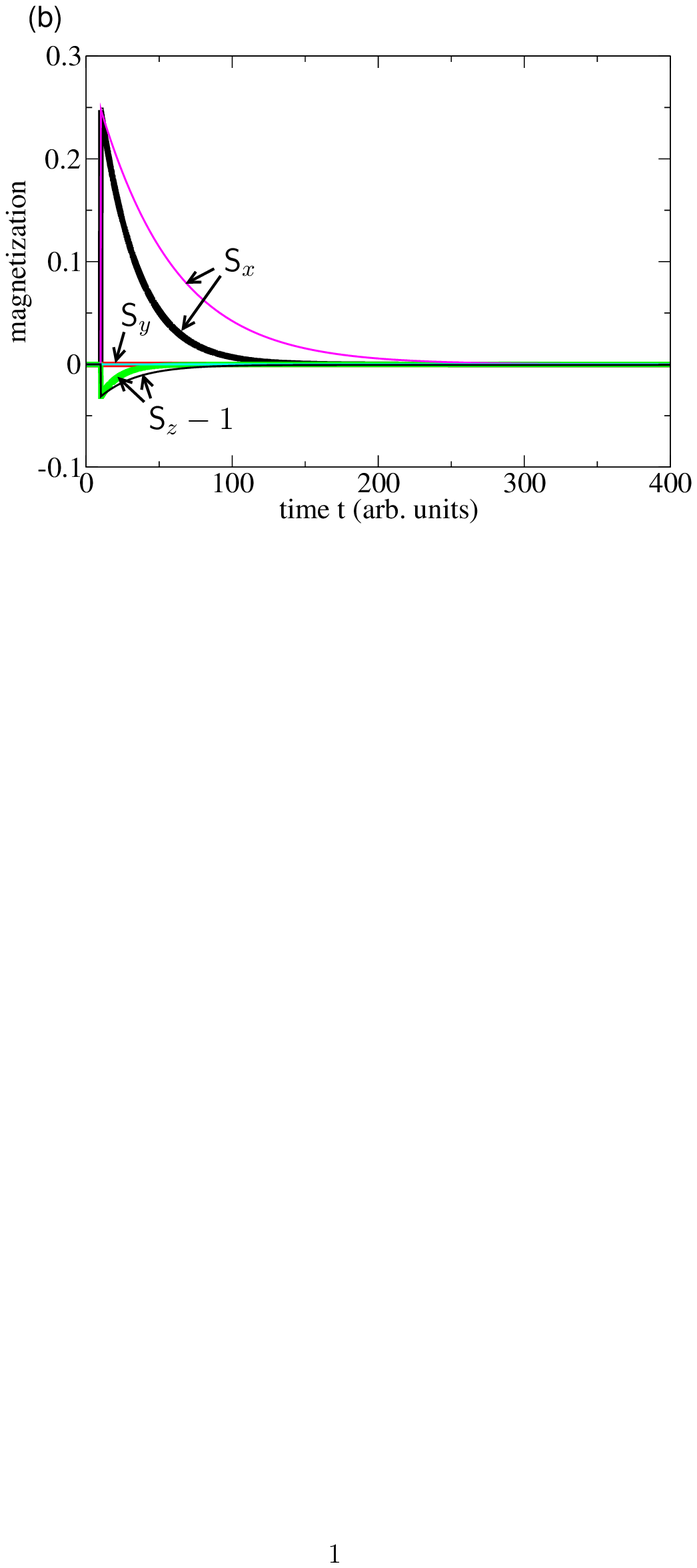}
  \end{center}
  \caption {(color online) Overdamped relaxation of magnetization after a 
gaussian field pulse. ${\sf S}_\alpha$, $\alpha \in\{x,y,z\}$ corresponds to 
classical (bold lines) resp. to quantum mechanical trajectories (thin lines). 
(a) $D_z = 0$ and $\mu_SB_z = 0.1$, (b) $D_z = 0.1$ and $\mu_SB_z = 0$. 
($T_W = 0.02$, $t_0 = 10$, $\mu_SB_0^x = 25.27$, and $\lambda = 0.2$)}     
  \label{f:pic3}
\end{figure}

In the case of the quadratic Hamiltonain with anisotropy we have the additional
term \cite{wieserPRB11} $\mathrm{i}D_z/\hbar\,(\hat{S}_n^x,\hat{S}_n^y, 0)$ 
which commutes with $ \mathbf{\hat{S}}$. Therefore the deviation between 
classical and quantum trajectory does not come from this term. The question is 
whether this correction comes from the precessional term 
$-\mathrm{i}/\hbar[\mathbf{\hat{S}}_n ; \hat{\mathrm{H}}]$ only or whether the 
relaxation term $\lambda/\hbar[\mathbf{\hat{S}} ; [\mathbf{\hat{S}}; 
\hat{\mathrm{H}} ]]$ also leads to a correction. To answer this question and to 
clarify the effect of the damping we compare the trajectories in the overdamped
limit $(\lambda \gg 1)$. Here we assume that the damping dominates the dynamics 
and skip the precessional terms: the overdamped TDSE is given by:
\begin{equation} \label{GisinOD}
\left(\frac{\mathrm{d}}{\mathrm{d}t} + \frac{\lambda}{\hbar} 
[\hat{\mathrm{H}} - \langle \hat{\mathrm{H}} \rangle]\right)|\psi(t)\rangle = 0
\;,
\end{equation}
and the overdamped Landau-Lifshitz equation by:
\begin{eqnarray} \label{LLOD}
\frac{\partial {\mathbf S}}{\partial t} = \frac{\lambda}{\mu_S}{\mathbf S}
\times \left({\mathbf S} \times {\mathbf H}_{\mathrm{eff}} \right) \,.    
\end{eqnarray}
Fig.~\ref{f:pic3} shows the trajectories of the overdamped relaxation process 
after a field pulse excitation. As expected in case (i) $D_z = 0$, $B_z \neq 0$
we see a perfect agreement between the quantum mechanical and the classical 
curve. In case (ii) $D_z \neq 0$, $B_z = 0$ we find the deviation which means 
that the second commutator also produces a correction which modifies the 
correction which comes from the precession term. 

In summary I have shown that it is possible to derive the Landau-Lifshitz 
equation from the quantum mechanical time evolution of a wave function. This 
derivation reveals the underlying mathematics and assumptions. During the 
derivation we get different presentations in different physical pictures and 
descriptions. 

In quantum mechanics we can find behavior which cannot be described by 
classical physics like quantum tunneling. In a previous publication 
\cite{wieserPRB11} I have shown that quadratic or higher order Hamilton 
operators do not behave classical, meaning the Ehrenfest theorem does not hold 
in these cases. In this publication this concept has been used to proof the 
damping term. In the case of a linear Hamiltonian we see a perfect agreement of
classical physics and quantum mechanics, but for quadratic and higher order 
Hamiltonians a deviation appears, which comes from the damping term. Therefore, 
the described formalism gives us the possibility to compare the classical 
with the quantum spin dynamics. 

The author wants to thank N. Mikuszeit and S. Krause for helpful discussions. 
This work has been supported by the Deutsche Forschungsgemeinschaft 
(SFB 668 B3) and the Hamburg Cluster of Excellence NANOSPINTRONICS.

\bibliography{Cite}
\end{document}